\documentclass[twoside,twocolumn,9pt]{article}
\usepackage{extsizes}
\usepackage[super,comma,sort&compress]{natbib}
\makeatletter
\newcommand*{\citenst}[2][]{%
  \begingroup
  \let\NAT@mbox=\mbox
  \let\@cite\NAT@citenum
  \let\NAT@space\NAT@spacechar
  \let\NAT@super@kern\relax
  \renewcommand\NAT@open{[}%
  \renewcommand\NAT@close{]}%
  \citealp[#1]{#2}%
  \endgroup
}
\makeatother
\usepackage{filecontents}
\usepackage[version=3]{mhchem}
\usepackage[left=1.5cm, right=1.5cm, top=1.785cm, bottom=2.0cm]{geometry}
\usepackage{balance}
\usepackage{mathptmx}
\usepackage{multirow}
\usepackage{sectsty}
\usepackage{graphicx} 
\usepackage{lastpage}
\usepackage[format=plain,justification=justified,singlelinecheck=false,font={stretch=1.125,small,sf},labelfont=bf,labelsep=space]{caption}
\usepackage{float}
\usepackage{fancyhdr}
\usepackage{fnpos}
\usepackage[english]{babel}
\addto{\captionsenglish}{%
}
\usepackage{array}
\usepackage{droidsans}
\usepackage{charter}
\usepackage[T1]{fontenc}
\usepackage[usenames,dvipsnames]{xcolor}
\usepackage{setspace}
\usepackage[compact]{titlesec}
\usepackage{hyperref}

\definecolor{cream}{RGB}{222,217,201}

\begin{document}

\pagestyle{fancy}
\thispagestyle{plain}
\fancypagestyle{plain}{
\renewcommand{\headrulewidth}{0pt}
}

\makeFNbottom
\makeatletter
\renewcommand\LARGE{\@setfontsize\LARGE{15pt}{17}}
\renewcommand\Large{\@setfontsize\Large{12pt}{14}}
\renewcommand\large{\@setfontsize\large{10pt}{12}}
\renewcommand\footnotesize{\@setfontsize\footnotesize{7pt}{10}}
\makeatother

\renewcommand{\thefootnote}{\fnsymbol{footnote}}
\renewcommand\footnoterule{\vspace*{1pt}%
\color{cream}\hrule width 3.5in height 0.4pt \color{black}\vspace*{5pt}} 
\setcounter{secnumdepth}{5}

\makeatletter 
\renewcommand\@biblabel[1]{#1}            
\renewcommand\@makefntext[1]%
{\noindent\makebox[0pt][r]{\@thefnmark\,}#1}
\makeatother 
\renewcommand{\figurename}{\small{Fig.}~}
\sectionfont{\sffamily\Large}
\subsectionfont{\normalsize}
\subsubsectionfont{\bf}
\setstretch{1.125}
\setlength{\skip\footins}{0.8cm}
\setlength{\footnotesep}{0.25cm}
\setlength{\jot}{10pt}
\titlespacing*{\section}{0pt}{4pt}{4pt}
\titlespacing*{\subsection}{0pt}{15pt}{1pt}

\fancyfoot{}
\fancyfoot[LO,RE]{\vspace{-7.1pt}\includegraphics[height=9pt]{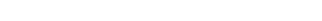}}
\fancyfoot[CO]{\vspace{-7.1pt}\hspace{13.2cm}}
\fancyfoot[CE]{\vspace{-7.2pt}\hspace{-14.2cm}}
\fancyfoot[RO]{\footnotesize{\sffamily{1--\pageref{LastPage} ~\textbar  \hspace{2pt}\thepage}}}
\fancyfoot[LE]{\footnotesize{\sffamily{\thepage~\textbar 1--\pageref{LastPage}}}}
\fancyhead{}
\renewcommand{\headrulewidth}{0pt} 
\renewcommand{\footrulewidth}{0pt}
\setlength{\arrayrulewidth}{1pt}
\setlength{\columnsep}{6.5mm}
\setlength\bibsep{1pt}

\makeatletter 
\newlength{\figrulesep} 
\setlength{\figrulesep}{0.5\textfloatsep} 

\newcommand{\topfigrule}{\vspace*{-1pt}%
\noindent{\color{cream}\rule[-\figrulesep]{\columnwidth}{1.5pt}} }

\newcommand{\botfigrule}{\vspace*{-2pt}%
\noindent{\color{cream}\rule[\figrulesep]{\columnwidth}{1.5pt}} }

\newcommand{\dblfigrule}{\vspace*{-1pt}%
\noindent{\color{cream}\rule[-\figrulesep]{\textwidth}{1.5pt}} }

\makeatother
\renewenvironment{abstract}
 {\small
  \begin{center}
  \bfseries \vspace{-.5em}\vspace{0pt}
  \end{center}
  \list{}{%
    \setlength{\leftmargin}{25mm}
    \setlength{\rightmargin}{\leftmargin}%
  }%
  \item\relax}
 {\endlist}

\twocolumn[
\begin{@twocolumnfalse}

\begin{center}
 \noindent\LARGE{\textbf{Exploring Noncollinear Magnetic Energy Landscapes\\ with Bayesian Optimization}} \\
\vspace{0.3cm} 

 \noindent\large{Jakob Baumsteiger,$^{\ast}$\textit{$^{ab}$} Lorenzo Celiberti,\textit{$^{b}$} Patrick Rinke,\textit{$^{cdef}$}\\ Milica Todorovi\'c,\textit{$^{g}$} and Cesare Franchini\textit{$^{ab}$}} \\
 \end{center}

\begin{abstract}
 \noindent\normalsize{The investigation of magnetic energy landscapes and the search for ground states of magnetic materials using \emph{ab initio} methods like density functional theory (DFT) is a 
challenging task. Complex interactions, such as superexchange and spin-orbit coupling, make these calculations computationally expensive and often lead to non-trivial energy landscapes. Consequently, a comprehensive and systematic investigation of large magnetic configuration spaces is often impractical.
We approach this problem by utilizing Bayesian Optimization, an active machine learning scheme that has proven to be efficient in modeling unknown functions and finding global minima. Using this approach we can obtain the magnetic contribution to the energy as a function of one or more spin canting angles with relatively small numbers of DFT calculations. To assess the capabilities and the efficiency of the approach we investigate the noncollinear magnetic energy landscapes of selected materials containing 3$d$, 5$d$ and 5$f$ magnetic ions: \ce{Ba3MnNb2O9}, \ce{LaMn2Si2}, $\beta$-\ce{MnO2}, \ce{Sr2IrO4}, \ce{UO2} and \ce{Ba2NaOsO6}. 
By comparing our results to previous ab initio studies that followed more conventional approaches, we observe significant improvements in efficiency.}
\end{abstract}
\end{@twocolumnfalse} \vspace{0.6cm}
  ]

\renewcommand*\rmdefault{bch}\normalfont\upshape
\rmfamily
\section*{}
\vspace{-1cm}


\footnotetext{\textit{$^{a}$~Department of Physics and Astronomy 'Augusto Righi', Alma Mater Studiorum - Univeristà di Bologna, 40127 Bologna, Italy. E-mail: jakob.baumsteiger2@unibo.it}}
\footnotetext{\textit{$^{b}$~Faculty of Physics and Center for Computational Materials Science, University of Vienna, 1090 Vienna, Austria. }}
\footnotetext{\textit{$^{c}$~Physics Department, TUM School of Natural Sciences, Technical University of Munich, Garching, Germany}}
\footnotetext{\textit{$^{d}$~Department of Applied Physics, Aalto University, FI-00076 Aalto, Finland. }}
\footnotetext{\textit{$^{e}$~Atomistic Modelling Center, Munich Data Science Institute, Technical University of Munich, Garching, Germany}}
\footnotetext{\textit{$^{f}$~Munich Center for Machine Learning (MCML)}}
\footnotetext{\textit{$^{g}$~Department of Mechanical and Materials Engineering, University of Turku, FI-20014 Turku, Finland. }}

\section{Introduction}
Magnetic materials are classified based on key properties like their magnetic ground state, exchange constants, magnetic anisotropy energy, and metastable or excited states.~\cite{Szilva} These properties are essential for understanding the origin of magnetic phases and for designing and optimizing materials for various technological applications.~\cite{moree2022,harris2019,tokura2020}
For simple materials, it is generally feasible to determine these properties through a combination of symmetry analysis and computational simulations. Approaches that aim to identify magnetic ground states typically begin by proposing several possible magnetic configurations. Ab initio methods, such as density functional theory (DFT), are then used to determine which configurations have the lowest energy.~\cite{paul2015,spivsak2000}
Alternative approaches can be employed, which rely on partially exploring the magnetic energy surface with respect to a single parameter, typically a canting angle. In such cases, the ground state can be estimated by performing a series of ab initio calculations for plausible values of the undetermined parameter.~\cite{lee2014,vzivkovic2022} The results can be fitted to a model Hamiltonian to derive magnetic exchange parameters.~\cite{Liu2015,dudarev2019,Mosca2021}
However, these simple approaches have very limited scope, and often break down when it comes to more complex magnetic materials, such as frustrated magnets or materials that show strong spin-orbit coupling. In such complex cases, it is very difficult to determine good guesses of the noncollinear ground states.
Moreover, when more than one parameter needs to be determined, traditional approaches to sample the relevant configuration space require a significantly higher number of ab initio calculations, thereby increasing computational time.
It is also important to note that complex magnetic materials require equally complex model Hamiltonians for an accurate description, making it generally unfeasible to perform a sufficient number of ab initio calculations for a successful fit.

In recent years, solutions to these problems have been attempted.
Payne \emph{et al.} used the meta heuristic firefly algorithm~\cite{payne2018} and Zheng and Zhang genetic evolution~\cite{zheng2021} to identify noncollinear ground states. Although successful, both attempts still require hundreds of DFT calculations per material to converge to the final result.
An approach by Huebsch \emph{et al.} aims to find magnetic ground states with the help of a basis set of magnetic structures determined by a cluster-multipole expansion, a theoretical approach based on magnetic point group analysis.~\cite{huebsch2021}
However, magnetic ground states are often unknown linear combinations of these basis configurations.

Although the aforementioned methods concentrate on identifying magnetic ground states, they overlook additional details about the magnetic energy landscape. Consequently, there is a need for an efficient approach to explore noncollinear magnetic energy landscapes to determine the ground state and other magnetic properties of complex magnetic materials. This work proposes a solution using a data-driven Bayesian optimization (BO) technique.

BO is a versatile machine learning tool that has found widespread applications across diverse domains ~\cite{koyama2017,yu2020,Todorovic2019,calandra2016,ilten2017,griffiths2020,rasmussen2006,Loefgren2022,Wu2024,Li2024}.
It excels in scenarios where evaluating the objective function is resource-demanding and proves particularly valuable for modeling complex, potentially multi-dimensional black box functions.
Here we use DFT-informed BO to explore noncollinear magnetic energy landscapes. Thereby we gain important insight into magnetic properties while uncovering metastable states and magnetic ground states at the same time.

We test our approach using the benchmark materials \ce{Ba3MnNb2O9}, \ce{LaMn2Si2}, $\beta$-\ce{MnO2}, \ce{Sr2IrO4}, \ce{UO2} and \ce{Ba2NaOsO6}. These materials represent a minimal dataset of materials characterized by different structural motifs, different magnetic orbitals, and different magnetic structures, as summarized in Table \ref{tab:dft_details}.
\begin{table*}[h]
\small
  \caption{\ Overview of the benchmark materials and their respective DFT calculation parameters}
  \label{tab:dft_details}
  \begin{tabular*}{\textwidth}{@{\extracolsep{\fill}}lllllll}
    \hline
    Material & Structure & Magnetic orbital & Magnetic ordering & k-mesh & Energy cutoff [eV] & Hubbard correction [eV] \\
    \hline
    \ce{Ba3MnNb2O9}   &  triple perovskite & 3d & $120^\circ$ Néel state & 5x5x6  & 500                    & $U_{eff}=3$ (ref. \citenst{lee2014}) \\
    \ce{LaMn2Si2}     & tetragonal & 3d & canted FM & 8x8x8  & 500                    & -                                                                   \\
    $\beta$-\ce{MnO2} & rutile & 3d & helical & 8x8x5  & 600                    & $U = 6.7$, $J = 1.2$ (ref. \citenst{tompsett2012}) \\
    \ce{Sr2IrO4}     & layered perovskite & 5d & canted AFM & 7x7x3  & 800                    & $U_{eff}=1.6$ (ref. \citenst{liu2016}) \\
    \ce{UO2}          & cubic & 5f & canted AFM & 6x6x6  & 700                    & $U_{eff}=3.46$ (ref. \citenst{dudarev2019}) \\
    \ce{Ba2NaOsO6}    & double perovskite & 5d & canted AFM & 6x6x6  & 600                    & $U_{eff}=3.4$ (ref. \citenst{Mosca2021}) \\
    \hline
  \end{tabular*}
\end{table*}

\section{Methods}
\subsection{Bayesian Optimization}
\label{sec:bo}
\begin{figure}[h!]
    \centering
    \includegraphics[width = 83mm]{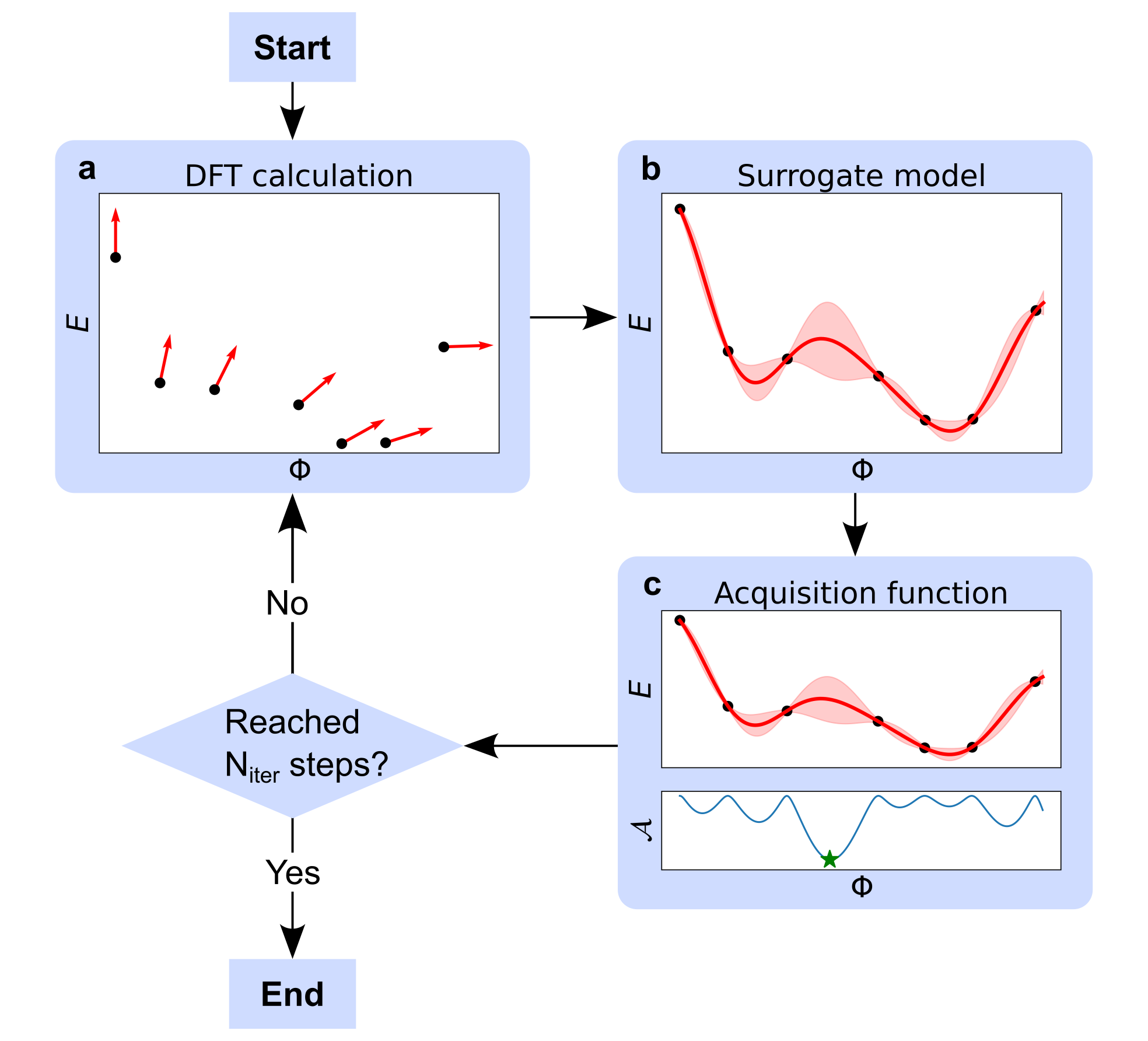}
    \caption{Workflow of the proposed approach. (a) At each iteration, a new DFT calculation is performed and the resulting energy value added to a data set. The canting angles $\Phi$ that correspond to the DFT calculations is indicated by red arrows.
    (b) Subsequently, a surrogate model $E_t (\Phi)$ (red) and the GPR native uncertainty $\sigma_t(\Phi)$ (red shade) are determined via GPR.
    (c) Using this information, the corresponding acquisition function $\mathcal{A}_t(\Phi)$ (blue) is calculated. The argmin (green star) of the acquisition function provides the parameters that specify the magnetic configuration for the next DFT calculation. After a predefined number of iterations $N_{\text{iter}}$, the optimization of the surrogate models ends.}
    \label{fig:flowchart}
\end{figure}

BO is an active machine learning method that iteratively evolves a regression surrogate model. Through an acquisition function BO determines the next data point at every iteration that is then added to the dataset. In this work, the dataset consists of magnetic configurations, defined by one canting angle $\Phi$ or several canting angles $\Vec{\Phi}$, and their corresponding DFT energies.
The main aspects of the BO workflow used here are shown in {\figurename \ref{fig:flowchart}}.

At each iteration step $t$, the BO algorithm constructs a surrogate model of the magnetic energy landscape $E_t(\Phi)$ and determines the corresponding statistical Bayesian uncertainty $\sigma_t(\Phi)$, utilizing the available data. The uncertainty $\sigma_t(\Phi)$ indicates the lack of confidence in the current model of the magnetic energy landscape that arises from the limited amount of available data and the noise of the DFT energies.
Typically, Gaussian process regression (GPR) is used for the generation of the surrogate model.\cite{brochu2010,rasmussen2006} The initial data set must include at least two data points to start the optimization procedure. We used two initial data points for the models involving one degree of freedom and four initial data points for all other models.

The configuration, which will be investigated in the next DFT calculation is strategically determined by utilizing an acquisition function $\mathcal{A}$, which usually depends on the current surrogate model and the Bayesian uncertainty. More precisely, the next DFT energy is determined for the magnetic configuration that minimizes the acquisition function.
Different acquisition functions have been proposed in the literature and acquisition function design is an active field of research. In this work we use the \textit{pure exploration} (EXP) acquisition function, which, in cour case, is expressed as
\begin{equation}
        \mathcal{A}^{\text{EXP}}_{t}(\Phi) = -\sigma_t(\Phi) \quad .
\end{equation}
$\mathcal{A}^{\text{EXP}}_{t}(\Phi)$ attains its minimum where the uncertainty $\sigma_t(\Phi)$ is highest, thereby promoting exploration. This is advantageous for efficiently exploring large phase spaces, which is the goal of our work.  

Once the BO algorithm terminates after completing $N_{\text{iter}}$ iteration steps, it is possible to evaluate whether the model has converged and, if so, determine the number of iterations required for convergence $N_{\text{iter}}^{\text{conv}}$ and the number of DFT calculations required for convergence $N_{\text{DFT}}^{\text{conv}}$. If convergence is not achieved after $N_{iter}$ iterations, the optimization can be resumed starting from the last iteration. Naturally, the total number of performed iterations must be larger than the number of iterations required for convergence $\left(N_{\text{iter}}^{conv}\right)$.

To ensure a high accuracy of the final models, we monitor the convergence of each surrogate model $E(\Vec{\Phi})$ using a validation set $S$, which consist of random magnetic configurations, parametrized by vectors containing all relevant canting angles $\Vec{\varphi}_i$ and the corresponding DFT energies $E_i$. We use the largest difference between any total energy within the validation set and the energy prediction of the surrogate model $(\Delta E)_{\text{max}}$ as a convergence measure:
\begin{equation}
    (\Delta E)_{\text{max}} = \max_{(\Vec{\varphi_i},E_i)\, \in\, S}  \, \left| E(\Vec{\varphi}) - E_i \right| 
\end{equation}
We consider convergence to be achieved, when $(\Delta E)_{\text{max}}$ falls below a threshold of 3\% of the energy range of the magnetic energy surface.
This threshold is chosen to ensure high enough accuracy of the models. While a more strict threshold would naturally result in a more precise model, the improvements could be considered negligibly small.
The validation sets contain 10, 20, 60, and 80 data points for the 1D, 2D, 3D, and 4D models, respectively. The size of the validation sets are a result of convergence tests that monitored the convergence of $N_{\text{DFT}}^{\text{conv}}$.

The concept and the mathematical foundation of BO are extensively described in the literature.\cite{rasmussen2006, garnett2023, brochu2010} In this work we used the Bayesian Optimization Structure Search (BOSS) \cite{Todorovic2019} python package (version 1.10.1.).

For the creation of the two- and three-dimensional models of the \ce{Ba3MnNb2O9} magnetic energy landscape, we exploit a mirror symmetry in the crystal structure to reduce the number of required calculations using a feature which is incorporated in BOSS. This feature allows us to to avoid redundant computations by adding not only the directly calculated DFT data point but also a second, symmetry-equivalent point to the data set.
We used the radial basis function kernel for the generation of almost all of our models. The only exception is our model of the magnetic energy landscape of \ce{UO2} that takes into account four individual degrees of freedom. In this case we used the Matérn kernel with $\nu=3/2$, since it outperformed the radial basis function kernel for this specific model in a kernel test not shown here. 

A key objective of this work is to assess the accuracy of the BO approach by comparing the resulting magnetic energy landscapes and ground state configurations with those reported in related previous studies.
To enable a meaningful comparison, we recreated the respective DFT setups used in the earlier studies. Consequently, the calculated DFT energies can be considered accurate within the context of the specific setups up to the numerical precision of our DFT calculations, which is $10^{-6}\,\mathrm{eV}$.
Since our aim is to recreate the features of the magnetic energy landscape within the framework determined by the previous studies, we set the GPR calculations in such a way that it can fit energy values with a noise characterized by a standard deviation of $10^{-6}\,\mathrm{eV}$.

\subsection{First principles calculations}
All DFT calculations were performed using the Vienna Ab initio Simulation Package (VASP),\cite{Kresse1996,Kresse1996_2,Hobbs2000} with spin-orbit coupling (SOC) incorporated. For all calculations we used the Perdew-Burke-Ernzerhof (PBE) exchange correlation functional \cite{pbe1996} and employed the projector augmented wave method \cite{kresse1999}.
To better describe the localized orbitals in strongly correlated materials, we primarily used the effective-$U$ method as outlined by Dudarev \emph{et al.}.\cite{dudarev1998}
An exception was made for $\beta$-\ce{MnO2}, for which we applied the method introduced by Liechtenstein \emph{et al.},\cite{liechtenstein1995} following the reasoning of Tompsett \emph{et al.}.\cite{tompsett2012}
The computational settings (k-points mesh, plane-wave cutoff, $U_{\text{eff}}$, or $U$ and $J$) are presented in Table \ref{tab:dft_details} for all materials.
To perform DFT calculations for specific non-collinear magnetic configurations, we used a constrained local magnetic moment approach.~\cite{ma2015, Liu2015}
The core concept of this method is to express the total energy $E$ as the sum of the DFT energy $E_0$ and a penalty term:
\begin{equation}
    E = E_0 + \sum_I \lambda \left[ \mathbf{M}_I - \hat{\mathbf{M}}_I^0(\hat{\mathbf{M}}_I^0 \cdot \mathbf{M}_I)\right]^2\, .\\
\end{equation}
Here $\hat{\mathbf{M}}_I^0$ is a normalized vector indicating the direction of the desired local magnetic moment at atom $I$ and $\mathbf{M}_I$ is the magnetic moment integrated over a sphere centered on atom $I$.
The parameter $\lambda$ determines the strength of the constraint.
It can be shown that the penalty term scales inversely with $\lambda$.\cite{ma2015} Consequently, increasing $\lambda$ to a sufficiently high value ensures that the contribution of the penalty term to the total energy becomes negligible. For most benchmark materials we set $\lambda=10$. For \ce{LaMn2Si2} and \ce{Ba2NaOsO6} we chose $\lambda=30$ and $\lambda=60$, respectively, in order to achieve sufficiently low penalty energies.

In the following, we showcase the capabilities of our approach using \ce{LaMn2Si2}, $\beta$-\ce{MnO2}, \ce{Ba2NaOsO6}, \ce{Sr2IrO4}, \ce{UO2} and \ce{Ba3MnNb2O9} as benchmark materials and monitor the computational efficiency.
We apply BO to generate up to four-dimensional models of the magnetic energy landscape.
Since the computational resources required to run the BO algorithm are negligible compared to the computational costs of DFT calculations, one can assume that the overall computational costs scale with the number of DFT calculations required for the convergence of the model.
The computation time for a single DFT calculation is heavily influenced by the computational system in use and the specific material being analyzed. Usually, materials of greater interest require more computation time since they are often more complex. In our case, the DFT calculations required between 30 minutes and four hours.

\section{Results}

\subsection{\ce{Ba3MnNb2O9}}
Geometrically frustrated materials received much attention recently because of the intriguing physics they exhibit and the complex theoretical challenges they present.~\cite{xu2023,amoroso2021,meschke2021} A common feature of frustrated magnetic materials is their unconventional and degenerate magnetic orders.
An interesting subgroup of frustrated materials are triangular lattice antiferromagnets, with \ce{Ba3MnNb2O9} as a representative example. At $0\,\mathrm{K}$ and in the absence of an external magnetic field, the magnetic ground state is a $120^\circ$ state, where all magnetic moments lie in the plane of the triangular lattice and the magnetic moments of each pair of neighboring lattice sites are oriented at $120^\circ$ angles relative to each other.~\cite{tian2014}
When an external magnetic field is applied parallel to the [001] direction, a collinear "up-up-down" (\textit{uud}) phase is stabilized within a finite range of magnetic field strengths.\cite{tian2014,lee2014}
We generate an one-, a two- and a three-dimensional model of the magnetic energy surface using BO, to directly compare the energy of the $120^\circ$ state with the \textit{uud} state and to search for possible metastable states. 

To determine an one-dimensional energy profile that connects the $120^\circ$ state and the \textit{uud} state, we define an out-of-plane canting angle $\Phi$ and quantify the energy gain as a function of $\Phi$ in the range between $0^\circ$ and $90^\circ$ using BO. Within this range, the one-dimensional magnetic energy surface lacks any symmetry that could be inferred from the crystal symmetry. Therefore we could not reduce the number of required DFT calculations by exploiting a given symmetry as described in Sec~\ref{sec:bo}.
The convergence criterion introduced in Sec~\ref{sec:bo} is reached after performing four DFT calculations.
The model and the four data points that were used for the generation of the model are displayed in {\figurename \ref{fig:figure_1}}a.
A similar energy profile, based on ten DFT calculations and qualitatively matching our results was previously determined by Lee \emph{et al.}~\cite{lee2014}
To further investigate the phase space for potential stable or metastable states, including those with lower symmetry, we created higher-dimensional models of the magnetic energy landscape by introducing two and three independent canting angles, respectively.
For two independent angles, we coupled $\Phi_2$ and $\Phi_3$ as defined in the inset of {\figurename \ref{fig:figure_1}}a and added $\Phi_1$ as a second independent canting angle. For three independent angles, we used all three canting angles as independent degrees of freedom. To account for both ferromagnetic and ferrimagnetic states, we expanded the search bounds of all angles to cover the range between $-90^\circ$ and $90^\circ$. As a consequence, the resulting models show a point symmetry that can be defined as $E(\Vec{\Phi})=E(-\Vec{\Phi})$, where $\Vec{\Phi}$ is a vector that contains the two or three canting angles, respectively. This symmetry follows from a mirror symmetry in the crystal structure.
The comparison of all three models is presented in {\figurename \ref{fig:figure_1}}a. As shown in the figure, all three models are in good agreement. {\figurename \ref{fig:figure_1}}b depicts a two-dimensional slice of the full three-dimensional model, which is defined by the constraint $\Phi_2=\Phi_3$. The three-dimensional model indicates that the $120^\circ$ state is the only stable or metastable configuration in the explored phase space.

\subsection{\ce{LaMn2Si2}}
\begin{figure*}[h!]
    \centering
    \includegraphics[width=171mm]{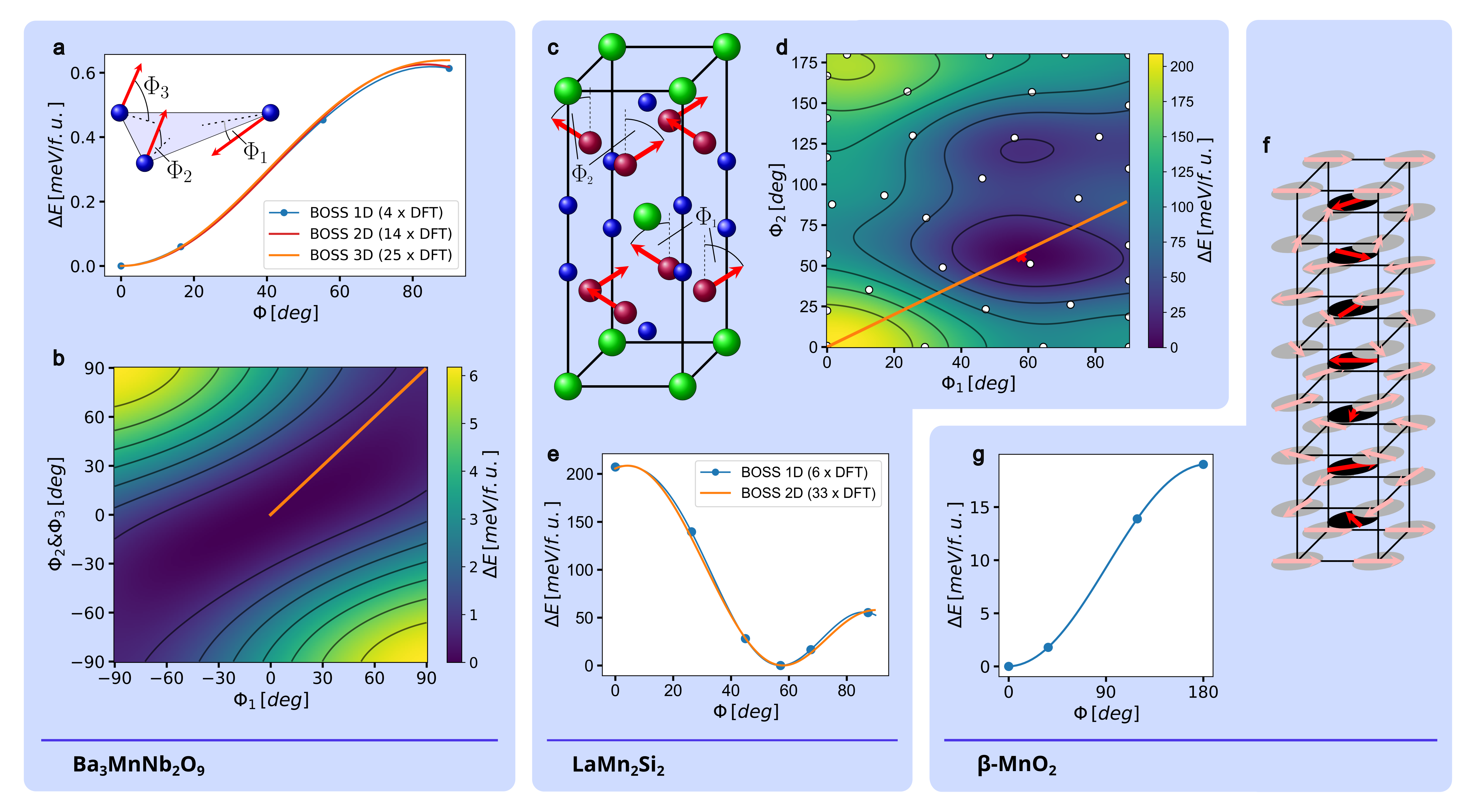}
    \caption{The definition of the canting angles used for the triangular lattice \ce{Ba3MnNb2O9} is shown in the inset of (a). In (b), a slice of the magnetic energy landscape of \ce{Ba3MnNb2O9} that was generated by defining one independent canting angle per magnetic site, is shown. The slice is defined by the constraint $\Phi_2=\Phi_3$. The comparison between a one-dimensional, a two-dimensional and a three-dimensional model of the magnetic energy landscape is shown in (a).
    The crystal structure of \ce{LaMn2Si2} and the used definition of the canting angles $\Phi_1$ and $\Phi_2$ are shown in (c). The magnetic energy landscapes that is spanned by the two canting angles, is depicted in (d).
    In (e), the energy profile corresponding to a canting angle $\Phi$ which is defined by the constraint $\Phi=\Phi_1=\Phi_2$, is shown. The results of the DFT calculations used for the creation of the model are depicted as blue points. The model is compared to the model from (d). The line that correspond to the definition of $\Phi$ is marked in (d) as an orange line.
    The reduced crystal structure of $\beta$-\ce{MnO2} is depicted in (f). Shown are also the magnetic moments of Manganese as proposed by Yoshimori.\cite{yoshimori1959} The body-centered Manganese atoms are displayed in solid colors. The effect on the total energy of rotating the canting angles of the body-centered Manganese atoms in plane by up to $180^\circ$, determined by BO is shown in (g).}
    \label{fig:figure_1}
\end{figure*}

\ce{RMn2X2} compounds, where R is a rare earth metal and X is \ce{Si} or \ce{Ge}, are  interesting due to their rich magnetic phase diagram.\cite{korotin2017,nowik1995,gerasimov2007} \ce{LaMn2Si2} is a \ce{RMn2X2} compound, that exhibits a canted ferromagnetic ground state below the Curie temperature.\cite{venturini1994,hofmann1997} In this state, the magnetic moments of the manganese atoms are aligned ferromagnetically along the [001], and antiferromagnetically along the [100] direction. The magnetic moments are canted with respect to the [001] direction by the canting angles $\Phi_1$ and $\Phi_2$ (see {\figurename \ref{fig:figure_1}}c). We generate an one- and a two-dimensional model of \ce{Sr2IrO4}'s magnetic energy surface to determine the canting angles in the magnetic ground state and to search for possible metastable states.

Using BO we obtain the energy difference as a function of the canting angle $\Phi$ defined by the constraint $\Phi=\Phi_1=\Phi_2$ in the range between $0^\circ$ and $90^\circ$. This range encompasses all possible magnetic configurations achievable by varying $\Phi$ due to the crystal's symmetry. The surrogate model is fully converged after the execution of seven DFT calculations within the BO framework and exhibits a global minimum that corresponds to a canting angle of $55.8^\circ$ (see {\figurename \ref{fig:figure_1}}e).
This finding agrees well with previous DFT studies, which estimated the canting angle to be between $53^\circ$ and $57.7^\circ$.\cite{dinapoli2004,korotin2017}

All \ce{Mn} atoms lie within two distinct lattice planes of the magnetic unit cell. To take into account additional magnetic configurations, particularly those that are antiferromagnetic along the [001] direction, we expand the magnetic energy landscape using BO by decoupling the canting angles of both planes and varying $\Phi_2$ between $0^\circ$ and $180^\circ$.
The resulting model is shown in {\figurename \ref{fig:figure_1}}d. It is converged after 32 DFT calculations. In addition to the global minimum, which coincides with the known magnetic ground state of \ce{LaMn2Si2} (see {\figurename \ref{fig:figure_1}}d and e), a local minimum emerges at $\Phi_1 = 59^\circ$ and $\Phi_2 = 121^\circ$, which indicates a metastable magnetic state. To the best of our knowledge, this state has not been reported yet as a stable state of \ce{LaMn2Si2}. The energy difference between both minima is $27.1\,\mathrm{meV/f.u.}$ and the difference between the global minimum and the saddle point between both minima amounts to $40.7\,\mathrm{meV/f.u.}$

\subsection{$\beta$-\ce{MnO2}}
Materials with helical spin structures are of great interest due to their compelling physics and possible spintronics applications. ~\cite{yokouchi2020,kurumaji2020,muehlbauer2009,koralek2009}
The first discovery of a helical magnetic phase was made in 1959 by Akio Yoshimori, who identified a helical structure in $\beta$-\ce{MnO2} through his analysis of neutron diffraction data.\cite{yoshimori1959}. Since then, $\beta$-\ce{MnO2} remained a prototypical example of helical magnets.
We employ BO to explore a one-dimensional configuration space that encompasses two helical states that were historically considered candidates for the magnetic ground state of $\beta$-\ce{MnO2}. To introduce the two states, we first provide a brief historical overview. Subsequently, we perform BO and generate a one-dimensional energy profile to determine whether one of the two states or an intermediate state is the energetic ground state.

As Yoshimori found out, the corner manganese atoms and the body-center cations form two distinct sublattices in the helical phase. The magnetic moments of both sublattices screw along the c-direction with a wavelength of 7c/2, so that the pattern repeats after seven unit cells (see {\figurename \ref{fig:figure_1}}f). All magnetic moments lie within the (001) plane. Later a magnetic X-ray diffraction experiment \cite{sato2001} and neutron diffraction experiments \cite{regulski2003,regulski2004} found that the wavelength of the helical structure is about 4\% smaller than 7c/2, which makes it incommensurate.
According to Yoshimori's model, the helical phase in $\beta$-\ce{MnO2} is stable only if the interaction between each corner atom and its neighboring body-centered atom is antiferromagnetic, eventually resulting in a $128.6^\circ$ tilt between each manganese atom and the two closest atoms of the other sublattice.

In 1966, W.P. Osmond argued that the interaction between corner atoms and body-centered atoms should be ferromagnetic according to the Goodenough-Kanamori rules for superexchange.~\cite{osmond1966} Based on this assumption he proposed an alternative pattern for the spin helix. This pattern differs from Yoshimori's model by inversion of the three-dimensional magnetic moment vectors of all body-centered atoms. This inversion can alternatively be described as a revolution of the three dimensional magnetic moment vectors of all body-centered atoms by $180^\circ$ around the c-axis.

To directly compare the energies of the spin configuration proposed by Osmond with the one predicted by Yoshimori, we use BO to determine the total energy of $\beta-$\ce{MnO2} as a function of the revolution angle $\Phi$ in the range between $0^\circ$ and $180^\circ$. Thereby $\Phi=0^\circ$ corresponds to the model of Yoshimori and $\Phi=180^\circ$ corresponds to the model of Osmond. The surrogate model function converges after four DFT calculations (see {\figurename \ref{fig:figure_1}}g). It clearly shows that the spin structure predicted by Yoshimori is lower in energy than the one predicted by Osmond. This result is in line with magnetic X-ray diffraction data acquired later.~\cite{sato2001}

\subsection{\ce{Sr2IrO4}}

\begin{figure*}[h!]
    \centering
    \includegraphics[width=171mm]{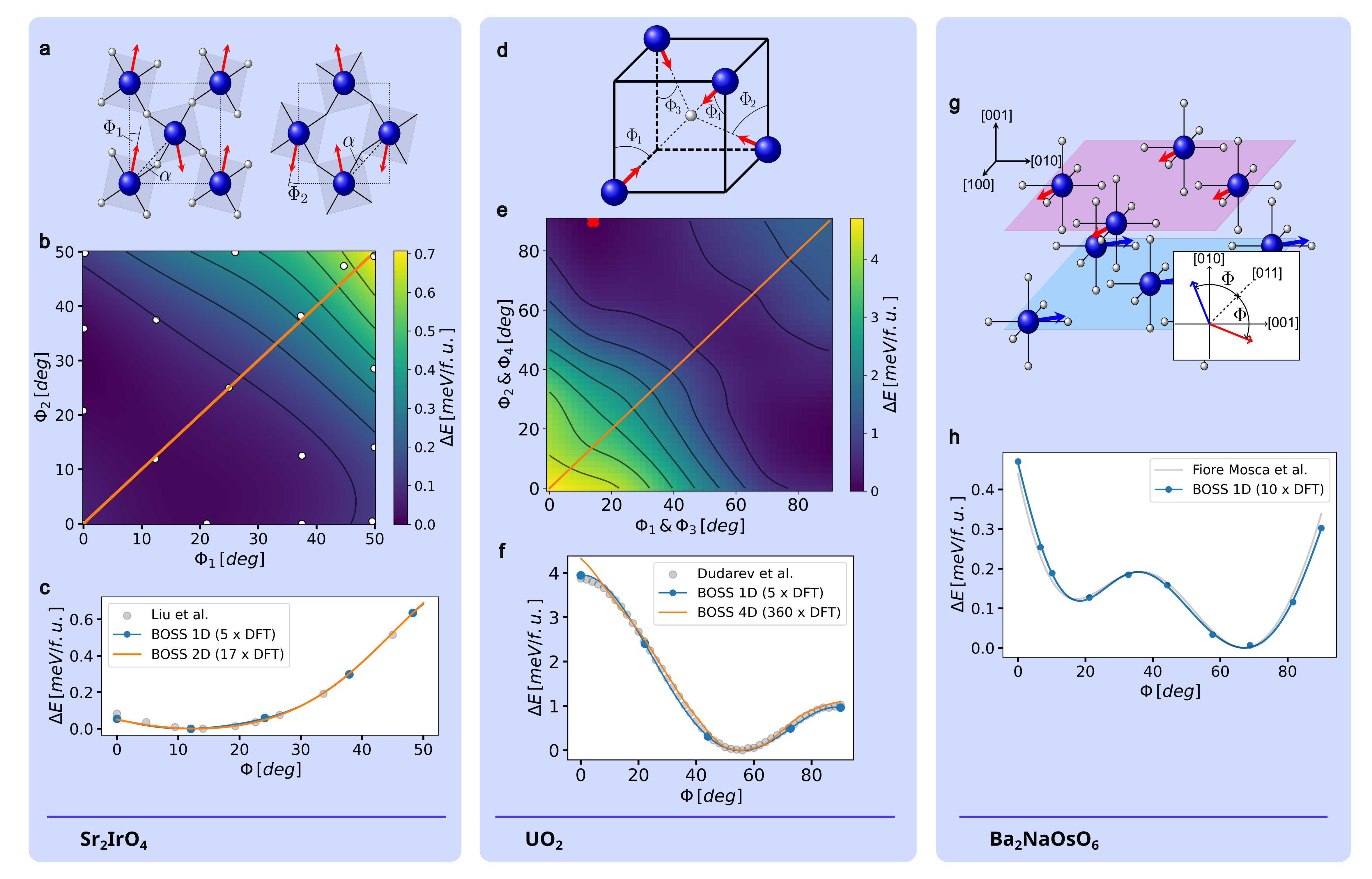}
    \caption{In (a), the crystal structures of subsequent lattice planes of \ce{Sr2IrO4} and the used definition of the canting angles $\Phi_1$ and $\Phi_2$ are given. The magnetic energy landscape spanned by the canting angles $\Phi_1$ and $\Phi_2$ is shown in (b). In (c), the two-dimensional model is compared to a one-dimensional model and to data from Liu \emph{et al.}.\cite{Liu2015} The crystal structure of \ce{UO2} as well as the used definition of the canting angles $\Phi_1$ to $\Phi_4$ are given in (d). (e) shows a slice of the four-dimensional model of the magnetic energy landscape obtained by using BO and all four canting angles independently. The slice is defined by the conditions $\Phi_1=\Phi_3$ and $\Phi_2=\Phi_4$. In (f) the four-dimensional model and a model obtained by a varying the four canting angles simultaneously are compared to data from Dudarev \emph{et al.}.\cite{dudarev2019} The reduced crystal structure of \ce{Ba2NaOsO6} is shown in (g). The definition of the canting angle $\Phi$ is given in the inset of the figure. The converged model of the BO energy profile is compared to data from Fiore Mosca \emph{et al.} in (h).}
    \label{fig:figure_2}
\end{figure*}
Strontium iridate (\ce{Sr2IrO4}) is a well-studied material that belongs to the Ruddelsden-Popper series. It was classified as a Dirac-Mott insulator.~\cite{kim2008,moon2008,Liu2021} Experimental results~\cite{Crawford1994} indicate that the \ce{IrO6} octahedra are tilted around the c-axis with an angle of $\alpha = 11.5^\circ$.
A DFT study by Liu \emph{et al.} \cite{Liu2015} found a similar rotation angle $\alpha$ of $13.2^\circ$.
Additionally, Liu \emph{et al.} determined a minimum energy canting angle of $\Phi = 12.3^\circ$ by calculating the energy of SIO for nine different canting angles $\Phi$ in the range between $0^\circ$ and $45^\circ$. We reproduce this result by generating a one-dimensional and a two-dimensional model of the magnetic energy surface using BO.

The canting angle $\Phi$ is defined by the condition $\Phi=\Phi_1=\Phi_2$, where $\Phi_1$ and $\Phi_2$ are illustrated in {\figurename \ref{fig:figure_2}}a. Therefore, $\Phi$ describes the common tilting of all magnetic moments from the crystallographic axis.
This result is in good agreement with experimental measurements ($\Phi=12.2^\circ$ \cite{boseggia2013}).
Using BO, we recreated the one-dimensional magnetic energy landscape within the interval from $\Phi=0^\circ$ to $\Phi=50^\circ$ (see {\figurename \ref{fig:figure_2}}b). The surrogate model converged after five DFT calculations.
The energy minimum of our model is reached at $\Phi=11.2^\circ$ which agrees well with the previous findings of Liu \emph{et al}.
Furthermore, we expanded the model of the magnetic energy landscape by decoupling the canting angles $\Phi_1$ and $\Phi_2$, considering magnetic configurations with different magnetic moment canting angles in neighboring planes. The resulting model, which converged after 17 DFT calculations, is shown in {\figurename \ref{fig:figure_2}}c. The comparison between this two-dimensional model and the data from Liu \emph{et al.} is shown in {\figurename \ref{fig:figure_2}}b.

\subsection{\ce{UO2}}
Most studies suggest that the magnetic ground state of uranium dioxide at $0\,\mathrm{K}$ is the so-called $3\mathbf{k}$ magnetic ordering, which is stabilized by strong SOC of the heavy element uranium.\cite{burlet1986,amoretti1989,blackburn2005,ikushima2001,wilkins2006,zhou2022}
In the $3\mathbf{k}$ state the magnetic moments of all four existing magnetic sites are tilted from the [001] direction by $54.7^\circ$. More details on the crystal structure of \ce{UO2} and the orientation of the magnetic moments in the $3\mathbf{k}$ crystal structure can be taken from {\figurename \ref{fig:figure_2}}d.
One of the computational studies that support the $3\mathbf{k}$ structure as the magnetic ground state of \ce{UO2} was conducted by Dudarev \emph{et al}.~\cite{dudarev2019}
Dudarev \emph{et al.} performed 46 constrained DFT+$U$ calculations with different values for a common canting angle $\Phi$ defined by $\Phi = \Phi_1=\Phi_2=\Phi_3=\Phi_4$ in the range between $0^\circ$ and $90^\circ$ to obtain a one-dimensional model of the magnetic energy landscape (see {\figurename \ref{fig:figure_2}}d for the definition of $\Phi_1$ to $\Phi_4$). The global minimum of their model coincides with the $3\mathbf{k}$ structure.

We could reproduce Dudarev's model with BO (see {\figurename \ref{fig:figure_2}}f)by generating an one-dimensional model with BO. The surrogate model converged after five DFT calculations.
By using all four canting angles as independent input parameters for BO that range between $0^\circ$ and $90^\circ$, we receive a full four-dimensional model of the magnetic energy landscape of uranium dioxide, which converges after 360 DFT calculations. A two-dimensional slice defined by $\Phi_1 = \Phi_3$ and $\Phi_2 = \Phi_4$ is shown in {\figurename \ref{fig:figure_2}}e.
Figure \ref{fig:figure_2}f demonstrates that our model is in good agreement with the data of Dudarev \emph{et al.}. However, our model indicates that the 3\textbf{k} state is not a global energetic ground state, but rather a saddle point in the magnetic energy landscape. The global minimum of our model is located at $\Phi_1=\Phi_3 = 14^\circ$ and $\Phi_2=\Phi_4=90^\circ$. The energy difference between the global energy minimum and the 3\textbf{k} state amounts to $0.38~\text{meV/f.u.}$
This result calls for further theoretical and experimental analysis for elucidating the magnetic ground state of UO$_2$.

\subsection{\ce{Ba2NaOsO6}}
The results of an NMR study conducted by Lu et.al. \cite{lu2017} revealed two inequivalent magnetic sites in the $5d$ double perovskite \ce{Ba2NaOsO6}. At temperatures below 7\,K, the magnetic moments tilt away from the [110] axis by $67^\circ$ in opposite directions while remaining within the $xy$ plane. This results in a canted antiferromagnetic configuration, as illustrated in {\figurename \ref{fig:figure_2}}g, which shows the crystal structure and the definition of the canting angles.
Fiore Mosca \emph{et al.} \cite{Mosca2021} demonstrated that the canted ground state arises from Jahn-Teller distortions of the \ce{OsO6} octahedra.
They used 36 single constrained DFT+$U$ calculations to model the dependence of the total energy on the canting angle $\Phi$. Between $\Phi = 0^\circ$ and $\Phi = 90^\circ$ they identified two energy minima, with the global minimum at $\Phi = 67^\circ$ closely matching the experimental results.
Employing BO within the same range of $\Phi$ values, we successfully reproduced the result.
The position of the two minima as well as the full energy profile are correctly captured by the BO model (see {\figurename \ref{fig:figure_2}}h). The model converged after nine DFT calculations were performed.

\section{Discussion}
Using BO, we generated models of the magnetic energy landscapes for six benchmark materials, each representing different material classes and magnetic orders. In all cases, the models converged after a relatively small number of DFT calculations. Where data from previous DFT studies was available, our results matched those findings.
This success is not surprising, as BO models have no intrinsic bias and will invariably converge to the correct magnetic energy landscape within the framework of DFT after a sufficient number of calculations.
Table \ref{tab:discussion} provides a summary of the BO performance across all benchmark materials, along with notable new findings that are presented in this work.
\begin{table*}
\small
\caption{Summary of BO performance compared to data from similar DFT studies.
The table includes the dimensionality and search bounds of the models, the number of DFT calculations required for convergence, and the number of DFT calculations used to construct similar models in related studies.
For \ce{Ba3MnNb2O9} we exploited symmetry as discussed in Sec~\ref{sec:bo}. The surrogate models in 2 and 3 dimensions were therefore fitted with twice as many points.
Furthermore, we added new findings from our studies to the table}
\label{tab:discussion}
\begin{tabular*}{\textwidth}{@{\extracolsep{\fill}}lllllll}
\hline
Material          & Dim. & Search bounds        & $N_{\text{DFT}}^{\text{conv}}$ this work &    $\left(N_{\text{DFT}}^{\text{conv}}\right)^{1/d}$ & $N_{\text{DFT}}$ related work & New findings     \\ \hline
\ce{Ba3MnNb2O9}   & 1D         & [-90$^\circ$, 90$^\circ$]                    & 4   &  4        & 10 (ref. \citenst{lee2014})         & - \\
                  & 2D              & [-90$^\circ$, 90$^\circ$]x[-90$^\circ$, 90$^\circ$] & 14   & 3.74         & -               &     -              \\
                  & 3D              & [-90$^\circ$, 90$^\circ$]x[-90$^\circ$, 90$^\circ$]x[-90$^\circ$, 90$^\circ$] & 25  &  2.92         & -          &   -            \\
\ce{LaMn2Si2}     & 1D           & [0$^\circ$, 90$^\circ$]          & 6     &  6      & -               & - \\
                  & 2D              & [0$^\circ$, 90$^\circ$] x [0$^\circ$, 180$^\circ$]   & 33 & 5.74   & -      &  Potential metastable state   \\
$\beta$-\ce{MnO2} & 1D              & [0$^\circ$, 180$^\circ$]                     & 4    &   4       & -               &    -  \\
\ce{Sr2IrO4}      & 1D            & [0$^\circ$, 50$^\circ$]        & 5   &    5      & 9 (ref. \citenst{Liu2015})       & - \\
                  & 2D              & [0$^\circ$, 50$^\circ$]x[0$^\circ$, 50$^\circ$]   & 17  & 4.12          & -               & -   \\
\ce{UO2}          & 1D             & [0$^\circ$, 90$^\circ$]                      & 5 &  5    & 46 (ref. \citenst{dudarev2019})  & - \\
                  & 4D              & [0$^\circ$, 90$^\circ$]x[0$^\circ$, 90$^\circ$]x[0$^\circ$, 90$^\circ$]x[0$^\circ$, 90$^\circ$]   & 360  &  4.35        & -               &     Potential ground state              \\
\ce{Ba2NaOsO6}    & 1D              & [0$^\circ$, 90$^\circ$]       & 10  & 10    & 36 (ref. \citenst{Mosca2021})     &      -    \\ \hline
\end{tabular*}
\end{table*}

Naturally, the number of DFT calculations needed for convergence $N_{\text{DFT}}^{\text{conv}}$ is most significantly influenced by the dimensionality of the model.
The value of $N_{\text{DFT}}^{\text{conv}}$ of the one-dimensional models ranges between four (\ce{Ba3MnNb2O9}, $\beta$-\ce{MnO2}) and ten (\ce{Ba2NaOsO6}). Compared to the number of DFT calculations used for in the respective related study, the BO approach required between 44\% (\ce{Sr2IrO4}) and 89\% (\ce{UO2}) fewer DFT calculations.
The two-dimensional models converged earliest after 14 DFT calculations (\ce{Ba3MnNb2O9}) and latest after 33 DFT calculations (\ce{LaMn2Si2}).
The only three-dimensional model presented in this work (\ce{Ba3MnNb2O9}) converged after 25 DFT calculations and the only four-dimensional model (\ce{UO2}) converged after 360 calculations.

Performance differences between models of the same dimensionality appear to be caused primarily by variations in their complexity. 
For instance, the only one-dimensional model with two local minima (\ce{Ba2NaOsO6}) required ten DFT calculations to converge, compared to the four to six calculations needed in the other cases. A similar pattern can be found for the two-dimensional models. The model for \ce{LaMn2Si2}, which features two local minima, required 33 calculations to converge. In contrast, \ce{Ba3MnNb2O9} and \ce{Sr2IrO4}, each with a single local minimum within the search bounds, required 14 and 17 calculations, respectively.
In both instances, $N_{\text{DFT}}^{\text{conv}}$ approximately doubles when the model contains two minima.

When $N_{\text{DFT}}^{\text{conv}}$ DFT calculations are required to converge a one-dimensional model, one could expect that $\left( N_{\text{DFT}}^{\text{conv}}\right)^d$ calculations are required to converge a $d$-dimensional model of the same material. However, in all cases discussed in this work, the performance of the multi-dimensional BO models is better than this approximation, as one can see from the $\left(N_{\text{DFT}}^{\text{conv}}\right)^{1/d}$ values in table \ref{tab:discussion}.

As stated in Sec~\ref{sec:bo}, we have used validation sets to track convergence of the surrogate models. This enabled us to ensure the convergence of the models and to meaningfully assess the performance of BO.
However, this approach might not be reasonable for the use of BO in most practical applications, since it requires additional DFT calculations. Therefore we recommend using the maximal change of the surrogate model per iteration step as a convergence measure. This measure led to similar results for the benchmark cases presented in this work.

The two-dimensional model of the magnetic energy landscape of \ce{LaMn2Si2} revealed a local metastable state defined by $\Phi_1=59^\circ$ and $\Phi_2=121^\circ$ following the definition of $\Phi_1$ and $\Phi_2$ given in {\figurename \ref{fig:figure_1}}c). Due to the high energy difference between the local minimum and the energy barrier, the existence of the local minimum can not be explained as an artifact of an imperfect choice of k-mesh or energy cutoff of the plane-wave expansion. To the best of our knowledge, this metastable state has not been reported yet.

In the case of \ce{UO2} there is a discrepancy between the postulated magnetic ground state and the minimum of the BO magnetic energy landscape.
Although our models match the literature DFT data perfectly, they show that the energy is further reduced for several magnetic configurations with reduced symmetry by up to $0.38~\text{meV/f.u.}$ The lower energy values of these configurations are not artifacts of the BO approach, since they are established by explicit DFT calculations, that are computed with an accuracy threshold of $10^{-6}~\text{eV}$ for a supercell containing four formula units \ce{UO2}. However, as mentioned in Sec~\ref{sec:bo}, we adapted the DFT setup of Dudarev \emph{et al.}~\cite{dudarev2019} in order to make the results comparable. This setup, in particular the choice of k-mesh and energy cutoff, might affect the observed magnetic ground state, especially considering the relatively small energy differences at play. Additionally, the choice of exchange-correlation functional might lead to unphysical results. It has been reported that the choice of exchange-correlation functional affects the total energy of \ce{UO2} obtained by DFT calculations in such a manner that it can result in different predicted ground states.\cite{zhou2022} However, although most studies indicate that the $3\mathbf{k}$ state is the physical ground state of \ce{UO2}, it remains a complicated matter and still poses a subject of debate.\cite{laskowski2004,zhou2022}
If the minima of our models are a result of the used exchange-correlation functional, the choice of the $U_{\text{eff}}$ parameter or if they have a real physical root will have to be clarified in further studies.

Despite the mentioned open questions concerning interpretation, the benchmark cases \ce{Ba3MnNb2O9} and \ce{UO2} highlight a strength of BO. They show how the capability of generating large magnetic energy landscapes with several degrees of freedom can help to spot interesting physics and potential problems of theory that would otherwise be overlooked.

BO-guided explorations are easy to implement and can easily be adapted to new problems. Although not explicitly discussed in this work, BO can also be applied to 2D magnets and materials with defects, where approaches based on symmetry analysis fail due to the breaking of symmetry by the defect.
Just like the methods used in previous approaches by Payne \emph{et al.} \cite{payne2018} and Zheng and Zhang,\cite{zheng2021} BO is an active machine learning scheme.
An important advantage of using active machine learning schemes is that they do not require large training datasets that would be computationally expensive to produce.

In general our results show that BO can be used as a tool to efficiently explore noncollinear magnetic energy landscapes and find magnetic ground states within large spin configuration spaces. Thereby it can help to obtain exhaustive insights into the intricate interactions of complex magnetic materials.

\section{Conclusion}
In this work we introduced a new approach to explore magnetic energy surfaces and find magnetic ground states efficiently by combining BO with constrained DFT+$U$ calculations.
Our approach exploits the strength of BO to accurately model unknown functions with a minimal number of data acquisitions.
We validated our method using \ce{Ba3MnNb2O9}, \ce{LaMn2Si2}, $\beta$-\ce{MnO2}, \ce{Sr2IrO4}, \ce{UO2} and \ce{Ba2NaOsO6} as benchmark materials - six complex magnetic materials that represent various types of magnetic structures.
As a first test for every material we defined a collective canting angle for all magnetic moments and determined the material's energy as a function of the canting angle using our BO approach.
The resulting one-dimensional surrogate models converge earliest after four and latest after ten DFT calculations. The ground states of all models are perfectly in line with previous studies.
Comparisons with studies on \ce{Sr2IrO4}, \ce{UO2} and \ce{Ba2NaOsO6} that investigated the dependence of the materials energies on the same collective canting angles using more classical approaches underline the high efficiency of our approach. In order to achieve the same results we needed between $44\%$ and $89\%$ fewer DFT calculations than in the respective study.
The gain in efficiency allowed us to expand the models to up to four degrees of freedom by defining more than one independent canting angle, while maintaining reasonable computational costs. Thereby we were provided with further insights into the magnetic energy landscapes of the six materials. Notably, the models revealed previously unreported stable and metastable magnetic states in \ce{LaMn2Si2} and \ce{UO2}.

\section*{Author contributions}
C.F.: conceptualization, supervision, writing - review \& editing.
J.B.: conceptualization, methodology, software, writing - original draft. L.C.: methodology, writing - review \& editing. M.T.: conceptualization, review \& editing. P.R.: conceptualization, review \& editing. 

\section*{Conflicts of interest}
There are no conflicts to declare.

\section*{Data availability}
Data for this article, including the central VASP input files (INCAR), the structural data files (POSCAR) and all calculated DFT energy values are available at Zenodo at https://doi.org/10.5281/zenodo.14528337.

\section*{Acknowledgements}
This work was supported by Piano Nazionale Resistenza e Resilienza (PNRR) - Next Generation Europe. The calculations have been performed using the Vienna Scientific Cluster (VSC).


\balance

\bibliography{rsc}
\bibliographystyle{rsc}
\end{document}